\documentclass{emulateapj}

\usepackage{hyperref}
\usepackage{color}

\shorttitle{Multi-Wavelength Study of a $\delta$-spot}
\shortauthors{Jaeggli}

\begin{document}

\title{Multi-wavelength study of a delta-spot \\ I: A region of very strong, horizontal magnetic field}

\author{S.~A. Jaeggli\altaffilmark{1,2}}
\affil{NASA Goddard Space Flight Center, Solar Physics Laboratory, Code 671, Greenbelt, MD 20771, USA}
\email{sarah.jaeggli@nasa.gov}

\altaffiltext{1}{Visiting Student, National Solar Observatory at Sacramento Peak, NM.  The NSO is operated by AURA, Inc.\ under contract to the National Science Foundation.}
\altaffiltext{2}{NASA Postdoctoral Program Fellow}

\begin{abstract}
Active region NOAA 11035 appeared in December 2009, early in the new solar activity cycle.  This region achieved a delta sunspot ($\delta$-spot) configuration when parasitic flux emerged near the rotationally leading magnetic polarity and traveled through the penumbra of the largest sunspot in the group.  Both visible and infrared imaging spectropolarimetry of the magnetically sensitive Fe I line pairs at 6302 \AA\ and 15650 \AA\ show large Zeeman splitting in the penumbra between the parasitic umbra and the main sunspot umbra.  The polarized Stokes spectra in the strongest field region display anomalous profiles, and strong blueshifts are seen in an adjacent region.  Analysis of the profiles is carried out using a Milne-Eddington inversion code capable of fitting either a single magnetic component with stray light or two independent magnetic components to verify the field strength.  The inversion results show that the anomalous profiles cannot be produced by the combination of two profiles with moderate magnetic fields.  The largest field strengths are 3500-3800 G in close proximity to blueshifts as strong as 3.8 km s$^{-1}$.  The strong, nearly horizontal magnetic field seen near the polarity inversion line in this region is difficult to understand in the context of a standard model of sunspot magnetohydrostatic equilibrium.
\end{abstract}


\section{Introduction}
Sunspots which have umbras with opposite magnetic polarities contained within a common penumbra are classified as $\delta$-spots in the Hale or Mt. Wilson classification scheme for active regions \citep{kunzel65}.  It is observationally well established that large, complex sunspot groups with $\delta$ configurations are the most likely to produce extreme flare events \citep{zirin87}.  \citet{guo14} found that over 90\% of X-class flares during solar cycles 22 and 23 were associated with active regions having a $\delta$ configuration, while \citet{sammis00} showed that flare energy is well correlated with both configuration and the area of a sunspot group.  Understanding the physical properties of $\delta$-spots is of critical importance for the modeling and prediction of major flares.

The $\delta$-spot configuration was first described by \citet{kunzel60}.  Active regions with $\delta$-spots are often complex, and sometimes have magnetic fields with reversed polarity and high shear.  They can evolve rapidly, especially during flare production \citep{liu05}.  They often have regions of hydrogen-$\alpha$ emission associated with flares or the emergence of new magnetic flux \citep{zirin87}.  The penumbra in the vicinity of the $\delta$-umbra can become highly distorted, with filaments wrapped tangentially around sunspot umbrae, and they sometimes display rapid rotation \citep{jiang12}.

In a few early measurements with spectropolarimeters, \citet{tanaka91} and \citet{zirin93a,zirin93b} observed large Zeeman splitting of spectral lines at the polarity inversion line (PIL) between the main polarity of the host sunspot and the opposite polarity umbra.  \citet{tanaka91} measured a field strength of 3500-4300 G transverse to the line of sight in an elongated pore and nearby penumbra fibrils over a 3-day period, and measured 4130 G in a light bridge between the main umbra and a $\delta$-umbra in another active region.  The PIL was also the location of strong photospheric upflows of 4 km s$^{-1}$.  The observations of \citet{zirin93a,zirin93b} show narrow lanes of strong field, from 3100 to 3980 G, between the opposite polarities of the $\delta$-spots in several different sunspot groups.

While modern imaging spectropolarimetry and adaptive optics have made it possible to study the photospheric properties of an entire active region at high spatial resolution, published observations have not shown further evidence for very strong fields at the $\delta$-spot PIL.  Based on the inversion of Stokes profiles in a $\delta$-spot, \citet{lites95} found a region of enhanced magnetic fill fraction at the PIL.  They specifically note that there is no corresponding enhancement or inconsistency in the magnetic field strength measured in the $\delta$-spot.  Based on the same observation, \citet{martinez94} showed evidence for supersonic downflows of at least 14 km s$^{-1}$ at the $\delta$-spot PIL.

\begin{figure*}
	\begin{center}	
		\includegraphics[width=7.25in]{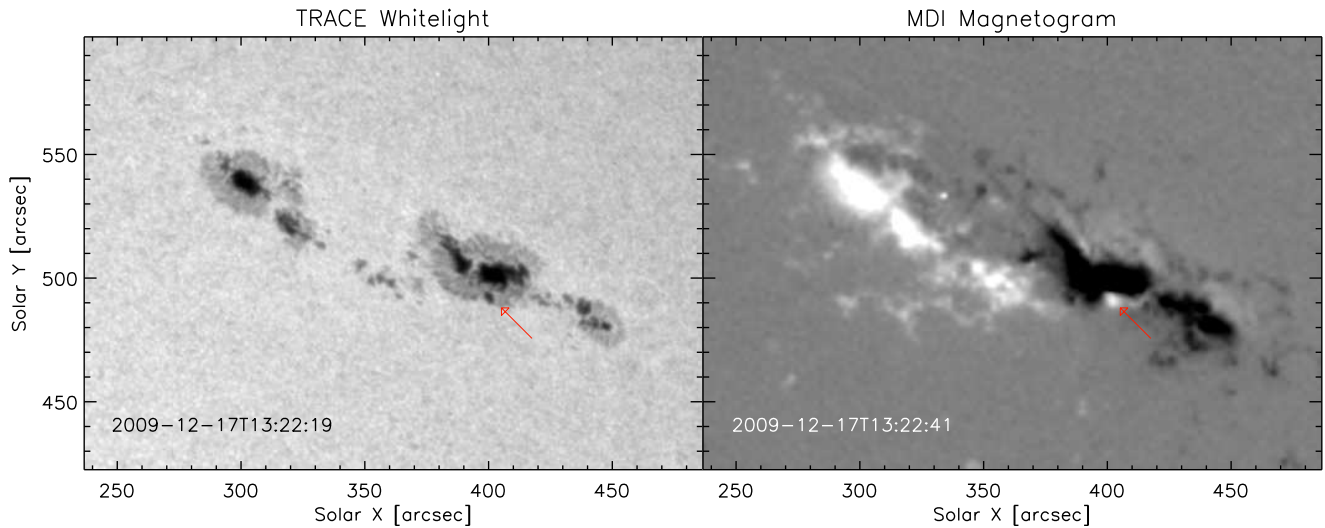}
	\end{center}
	\caption{TRACE whitelight image and MDI magnetogram near the time of the SOT/SP and DST/FIRS observations.  The arrow indicates the feature of interest in this study.}
	\label{fig:context}
\end{figure*}

More recently, \citet{balthasar14} showed the magnetic properties of a small opposite polarity pore in a bigger spot based on near-IR spectropolarimetry.  The magnetic field at the PIL in their active region does not show intensification of the transverse field, the field transitions smoothly from the main umbra into the opposite polarity patch and only small velocities are measured.  \citet{cristaldi14} also gave a recent analysis of a $\delta$-spot with high resolution spectropolarimetric data.  Intensification of transverse magnetic field is apparent at the PIL, although its strength only reaches a maximum of 2300 G (private communication).  The PIL is the site of both up and downflows along the line of sight of about 3 km s$^{-1}$.

Although very strong, horizontal magnetic fields in sunspots seem to be rare, such strong magnetic features are interesting because they provide a glimpse of the sunspot magnetic field under extreme conditions which may provide insight into more typical processes of penumbral formation, the forces which guide topological changes in active region magnetic fields, and conditions for reconnection.

A strong, compact magnetic feature was identified in the $\delta$-spot of NOAA 11035 based on observations taken two consecutive days with the Facility Infrared Spectropolarimeter (FIRS) on the Dunn Solar Telescope (DST).  \citet{jaeggli12} noted that this feature appeared to be in excess of 3500 G and was outside the main sunspot umbra, but it was not analyzed further.  In addition to FIRS, several space observatories tracked this active region throughout its formation, evolution, and eventual decay, including Hinode and the Transition Region and Coronal Explorer (TRACE) in addition to synoptic observations from the Solar and Heliospheric Observatory (SOHO).  The spectropolarimetric rasters of NOAA 11035 obtained with FIRS and the Hinode Spectropolarimeter on the Solar Optical Telescope (SOT/SP) are far more comprehensive than the observations presented in the work of \citet{tanaka91}, \citet{zirin93a,zirin93b}, and make it possible to study this very interesting strong field region in extreme detail.

The enhanced magnetic field is apparent in multiple observations of the region from December 17 to 18, but this paper (Paper I) considers a single set of observations from the Spectropolarimeter on Hinode's Solar Optical Telescope (SOT/SP) and ground-based observations from FIRS that were taken at approximately the same time on December 17th in order to verify the strength and direction of the extreme magnetic field.  The observations and data processing techniques are described in Sections \ref{sec:obs} and \ref{sec:data}.  A Milne-Eddington inversion technique is used to interpret the observations assuming a model with a single magnetic component (Section \ref{sec:anal1}).  This model is insufficient to describe the observed Stokes spectra in the region with the strongest field (Section \ref{sec:anal2}), and a Milne-Eddington inversion with two magnetic components is used to interpret these spectra (Section \ref{sec:anal3}).  A brief discussion of the results is given in Section \ref{sec:discuss}.  A subsequent paper (Paper II) will discuss results from the full time series of spectropolarimetric observations and additional space and ground-based imaging that provide insight into the evolutionary processes that produce this magnetic anomaly.

\section{Observations}\label{sec:obs}
The active region NOAA 11035 emerged near the central meridian on December 14, 2009, early in solar cycle 24.  The magnetic field of the region was highly sheared, and in accordance with Joy's law \citep{hale19}, it appeared at high latitude ($510''$ North of disk center) and had a large tilt with respect to the solar equator.  The region evolved rapidly over the next few days, producing two large sunspots and several smaller sunspots and pores, which decayed to a simple sunspot pair as the active region reached the limb on December 20.  During the growth of the active region, parasitic flux emerged into the rotationally leading polarity of NOAA 11035 and interacted with the main sunspot, producing a $\beta\delta$ magnetic configuration on December 16.  Figure \ref{fig:context} shows context images from the TRACE whitelight channel and a magnetogram from the Michelson-Doppler Imager on SOHO at the approximate time of the spectropolarimetric rasters taken with SOT/SP and FIRS when the main sunspot was at a heliocentric position of $\mu=0.754$.  A movie of the TRACE and MDI observations is available in the online material.  The parasitic polarity persisted throughout the interaction with the sunspot as it traveled through the southern sunspot penumbra, but it separated from the main sunspot and then abruptly vanished early December 19.

A spectropolarimetric raster of the lead sunspot in NOAA 11035 was taken with SOT/SP from 13:30:05 to 13:54:20 UT on December 17, 2009 using the fast map mode.  These observations cover a 2.39 \AA\ spectral window around 6302 \AA\ with a spectral resolution of 25 m\AA. This region of the spectrum includes a pair of Zeeman-split Fe I lines at 6301.5 and 6302.5 \AA\ with Land\'e factors of 1.5 and 2.5 respectively.  The fast map observations are binned by 2 spatially and spectrally, resulting in a spatial sampling of $0.30''\times0.32''$ while still covering a region of $151''\times162''$ \citep{lites13a}.

Observations with FIRS were taken from 13:44:51 to 14:08:46 UT on December 17, 2009.  FIRS is a dual-channel imaging spectropolarimeter capable of obtaining simultaneous measurements of the full-Stokes vector of visible (Fe I 6302 \AA) and infrared (He I 10830 \AA\ or Fe I 15650 \AA) magnetic field diagnostics on two separate detectors.  The efficiency of the raster observations is increased by the use of multiple slits.  However, this makes it necessary to use broad order-sorting filters paired with narrow-band dense wavelength division multiplexing (DWDM) filters in the optical path to prevent the spectra from adjacent slits from overlapping.  A pair of liquid crystal variable retarders in place immediately before the detector is used to modulate the polarization for each wavelength.  
The polarization is then analyzed with a Wollaston prism (a polarizing beam splitter) which produces two orthogonally polarized states which are vertically separated on the detector and are combined in post-processing to reduce spurious seeing effects.  See \citet{jaeggli10} and \citet{jaeggli11} for more instrument details.

For this observation the 6302 \AA\ and 15650 \AA\ channels were taken simultaneously to provide a robust measure of the photospheric magnetic field.   The SOT/SP observation at approximately the same time is free from seeing with a spatial resolution superior to the 6302 \AA\ channel of FIRS, so only the infrared data from FIRS is considered here.  The 15650 \AA\ channel of FIRS covers a bandpass of 10 \AA\ which contains a pair of Fe I lines at 15648.5 and 15652.9 \AA\ with large Land\'e factors of 3.0 and 1.67 respectively, which make them a highly sensitive tool for probing the photospheric magnetic field.  In addition to the Fe I lines, there are several strong OH lines in this bandpass which form only in cool sunspot umbrae.  They display the molecular Zeeman effect; two lines at 15651.874 and 15653.480 \AA\ become blended with the wings of the 15652.9 \AA\ line at moderate field strengths. 

During the observation, near diffraction-limited performance was achieved in the infrared (0.52$''$ at 15650 \AA) under good seeing conditions using the high-order adaptive optics system on the DST \citep{rimmele04}.  The FIRS slit unit was oriented perpendicular to the horizon so that atmospheric refraction was only in the dimension along the slit, making the IR and visible observations at each slit position truly simultaneous.  The spectrum at each polarization state is the coaddition of 3 exposures of 125 msec for the infrared and 2 exposures of 350 ms for the visible.  The science observations were followed immediately by darks and flat field observations of disk center.  To determine the spatial scale, observations of the line grid slide in place at the DST prime focus \citep{elmore92} were made at the beginning of the observing run on December 13.

The precise wavelengths and line parameters for the observed lines of significance are given in Table \ref{tab:atmlines} and \ref{tab:mollines}.  Values of log(gf) for the Fe I 15650 \AA\ lines were taken from \citet{borrero03} and parameters for the Fe I 6302 lines were taken from \citet{orozcosuarez10} which both fit for the log(gf) values using a two component quiet-Sun model.  Wavelengths of Fe I are as originally listed in \citet{nave94}.  Based on Figure 11 in \citet{berdyugina02} we have approximated a value of 0.15 for the effective Land\'e g of the OH lines, but for the purposes of the analysis this value appears to be too large.  The value listed in the table is used during the analysis (see more discussion on this in Section \ref{sec:anal1}).

\begin{deluxetable*}{rcccll}
	\tablecaption{Atomic Line Parameters\label{tab:atmlines}}
	\tablehead{\colhead{Wavelength [\AA]} & \colhead{Ion} & \colhead{Transition} & \colhead{J$_u$-J$_l$}& \colhead{$g_{eff}$} & \colhead{$\log{gf}$}}
	\startdata
		6301.5012 & Fe I & 3d$^6$ 4s 4p z $^5$P$^\circ$ - 3d$^6$ 4s 5s e $^5$D & 2-2 & 1.5 & -0.718 \\
		6302.4936 & Fe I & 3d$^6$ 4s 4p z $^5$P$^\circ$ - 3d$^6$ 4s 5s e $^5$D & 1-0 & 2.5 & -1.235 \\
		\hline
		15648.515 & Fe I & 3d$^6$ 4s 5s e $^7$D - 3d$^6$ 4s 5p n $^7$D$^\circ$ & 1-1 & 3.0 & -0.675 \\
		15652.874 & Fe I & 3d$^6$ 4s 4d f $^7$D - 3d$^6$ 4s 4f $^2$[7/2]$^\circ$ & 5-4 & 1.53 & -0.043
	\enddata
\end{deluxetable*}

\begin{deluxetable*}{rcccccll}
	\tablecaption{Molecular Line Parameters\label{tab:mollines}}
	\tablehead{\colhead{Wavelength [\AA]} & \colhead{Species} & \colhead{Branch} & \colhead{I$_u$-I$_l$} & \colhead{V$_u$-V$_l$} & \colhead{J$_u$-J$_l$} & \colhead{$g_{eff}$} & \colhead{$\log{gf}$}}
	\startdata
		15651.896 & OH & P & 1-1 & 3-1 & 5.5-6.5 & 0.1* & -5.132 \\
		15653.480 & OH & P & 1-1 & 3-1 & 5.5-6.5 & 0.1* & -5.132
	\enddata
	\tablenotetext{*}{Value assumed in this work.}
\end{deluxetable*}

\section{Data Reduction}\label{sec:data}
Reduced level 1 data from the SOT/SP were obtained from the Community Spectropolarimetric Analysis Center archive.  These data were already processed with the IDL SolarSoft routine \verb#SP_PREP# and were dark and flat corrected and corrected for orbital shifts.  The beams from the dual channel polarimetry were merged, the polarization states were demodulated and corrected for residual polarization cross-talk \citep{lites13a}.  The level 1 data was corrected for bit shifting (which can introduce a factor of 2 to the QUV components) and Stokes I, Q, U, and V were normalized to the average of the local quiet-Sun continuum level in Stokes I.

The reduced SOT/SP spectra show a standard deviation of between 0.4 and 0.1$\%$ (Stokes Q) of the continuum level in Stokes I.  The accuracy of the polarization calibration is thought to be comparable to the 0.1 \% photometric accuracy achieved in the normal map mode \citep{ichimoto08}.

The FIRS observations were processed using new techniques from those described in \citet{jaeggli12}.  All images were first dark subtracted using a dark image constructed by taking the mean of 64 frames.  The detector linearity curve in \citet{jaeggli11} was adjusted for the average dark level and then applied to the dark-subtracted image.  The geometric transformation from pixel coordinates to spatial and spectral coordinates was determined for each slit in each beam based on a 2D polynomial fit to the centers of spectral lines and grid lines in calibration images taken with the line grid.  This transformation was applied by 2D interpolation and resulted in a spectrum for each slit and beam which was coaligned with every other slit and beam, and had orthogonal spatial and spectral dimensions.  This geometrical transformation was based directly on the method used for spectra from the Interface Region Imaging Spectrograph \citep{depontieu14}.  The sequence of flat field images was averaged together and summed over the modulation states to produce a Stokes I frame.  This intermediate flat field was geometrically transformed and averaged along the spatial dimension over all slits to obtain the average spectral profile.  The two major Fe I lines were fitted and the resulting spectral profile of the lines was reversed though the geometric transformation and divided from the intermediate flat field to produce a final flat field without spectral lines, but still containing the detector structure, filter profiles, and intensity variation along each of the slits.

Scans were reconstructed from the science data by dark subtracting each frame and correcting for linearity.  The data was demodulated to obtain the full Stokes (I,Q,U,V) spectrum.  The flat field was divided from all four components and the geometric transform was performed, yielding $4\times2$ spectra from each slit and beam.  The two spectra from each beam were combined by addition in the case of Stokes I and by subtraction for the polarized components.

Following the reconstruction of the raster, the scan data cube was processed to remove residual flat field patterns and normalize the intensity to the local quiet-Sun continuum.  The FIRS infrared spectra suffer from low frequency optical fringes, due to the interference of light in the optical path, and high frequency digital fringes, from the detector electronics.  The fringe patterns evolve slowly with time, and fringes do not show up in the demodulated polarization states, which are differences of consecutive images a few seconds apart.  However, fringes can be strong in the intensity spectrum.  The flat fields used to correct the intensity in the scan are often 30 minutes apart and thermal drift of the fringes can amplify the pattern in the processed data.  The optical and digital fringe patterns were particularly strong in this data and proper removal was necessary for fitting the Zeeman-split line profiles in Stokes I.  Three strong frequencies were identified based on the power spectra of the original images.  Sine functions with the identified frequencies were simultaneously fit to the continuum regions of the spectrum at each position along the slit, with free parameters for amplitude and phase of each sine function.  The resulting fringe pattern was then divided from the entire spectrum.

After this step, instrumental polarization was measured and removed by the ad-hoc technique of \citet{kuhn94}, which assumes that polarization is exchanged primarily between the linearly and circularly polarized states and that, statistically for a sunspot, the Stokes Q and U profiles are symmetric, and the Stokes V profiles are anti-symmetric.  The polarization cross-talk was determined based on the main photospheric line (Fe I 15648.9 \AA) using spectra from all regions having sufficiently strong polarization signatures and low intensity (i.e. in the sunspot).

The resulting Stokes U spectra from FIRS show a standard deviation of less than 0.1$\%$ of the continuum level in Stokes I, although the errors do not follow typical photon counting statistics due to the detector non-linearity and the presence of digital fringes.

\section{Inversion with 1 magnetic component}\label{sec:anal1}
\subsection{Setup}
The magnetic field parameters were determined from the SOT/SP and DST/FIRS full-Stokes spectra using the Two Component Magneto-Optical (2CMO) Milne-Eddington inversion code.  The Milne-Eddington approximation simplifies the problem of polarized radiative transfer by decoupling the continuum emission and line absorption by using a linear source function and a constant line to continuum absorption ratio.  For the purposes of the inversion, the atomic parameters, velocity, and magnetic vector are assumed to be constant with depth in the line-forming region.

The 2CMO code is implemented in IDL and includes the ability to interactively invert profiles.  It uses a Levenberg-Marquardt least-squares minimization to fit the full-Stokes spectrum with Faraday-Voigt profiles approximated by the method of \citet{matta71}.  2CMO is based heavily on the formalism of \citet{jefferies89}, and the interested reader should refer to this paper for more details.  The code has been updated from the version used in \citet{jaeggli12} with an improved initialization for the magnetic field strength following the center of gravity technique of \citet{rees79}, and for the magnetic field azimuth and inclination angles following \citet{auer77}.  The previous version, with a one magnetic component and a stray light component, produced results nearly identical to the Milne-Eddington gRid Linear Inversion Network (MERLIN) code \citep{jaeggli12} which is the descendant of the code developed by \citet{skumanich87}.

\begin{figure*}
	\begin{center}
		\includegraphics[width=7in]{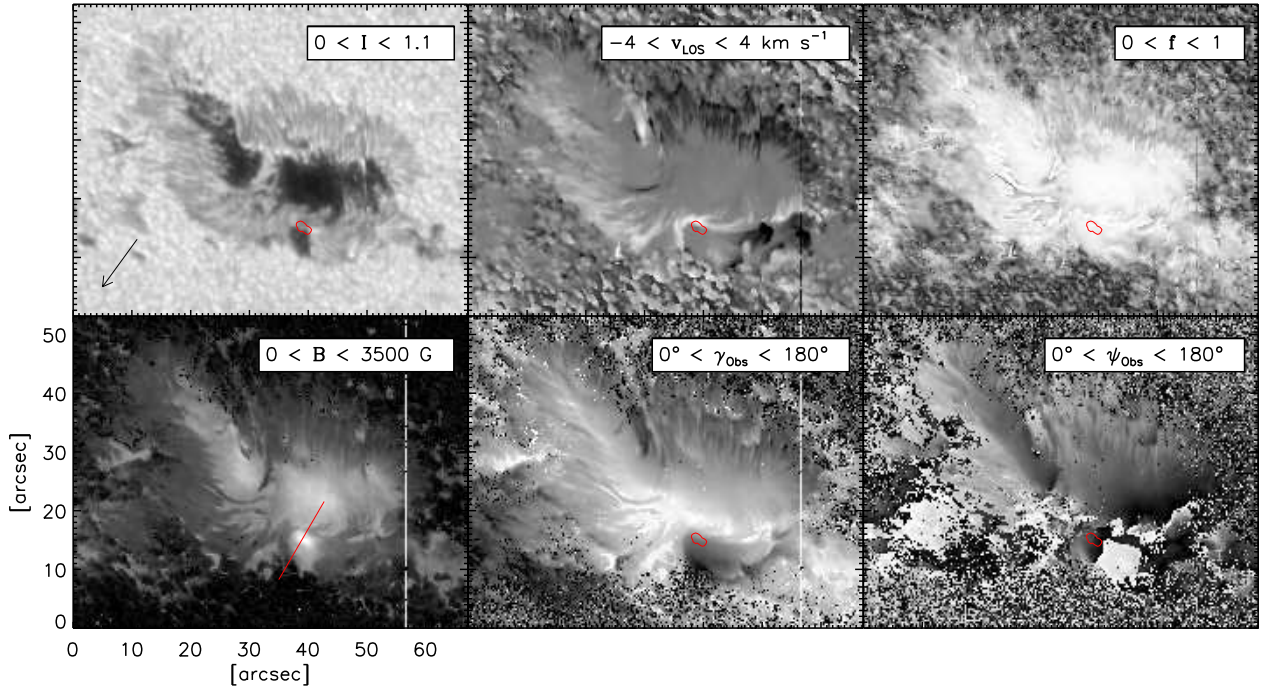}
	\end{center}
	\caption{Parameter maps from a sub-field of the SOT/SP 6302 \AA\ raster observation of NOAA 11035 taken 13:30:05 - 13:54:20 UT on Dec 17, 2009.  Continuum intensity (top left) was determined directly.  The other parameters are the result of inversion with the 2CMO 1M+SL model: line of sight velocity (top center), magnetic fill fraction (top right), magnetic field strength (bottom left), magnetic field inclination to the line of sight (bottom center), and the magnetic field azimuth in the plane of the sky (bottom right).}
	\label{fig:SOTSP_maps}
\end{figure*}

The inversion was performed for a sub-section of the full SOT/SP and FIRS field of view using a model with a single magnetic component and a stray light component (hereafter called 1M+SL).  The magnetic component includes eight free parameters:  the source function (B$_0$), the source function gradient (B$_1$), the Doppler width of the line ($\sigma$), the line damping parameter ($\alpha$), the line to continuum absorption ratio ($\eta_0$), the magnetic field strength (B), the magnetic field azimuth in the plane of the sky ($\psi_{Obs}$), and the magnetic field inclination to the observed line of sight ($\gamma_{Obs}$).  The stray light profile used by the inversion is an average of the quiet-Sun profiles within the field of view.  This does not require any prior knowledge about the stray light properties of the instrument, however the stray light component cannot be distinguished from an unresolved non-magnetic component blending within the pixel.  The stray light profile is allowed a free scaling parameter (1-f) and therefore the magnetic filling factor is (f).

Accurate inversion of the 15650 \AA\ Stokes profiles in a sunspot requires also fitting the lines of OH which blend with the wings of the Zeeman-split 15652.9 \AA\ Fe I line at umbral temperatures and field strengths.  In 2CMO, the line profile used for the OH lines is a Gaussian instead of a Voigt profile.  The OH lines are given separate parameters for line width ($\sigma_{OH}$) and line amplitude (f$_{OH}$), but it is assumed they experience the same magnetic field and velocity.

Line parameters used in the inversions are listed in Table \ref{tab:atmlines} and \ref{tab:mollines}.  The FIRS data were inverted and then the resulting parameter maps were shifted and rotated to match the SOT/SP orientation.

Monte Carlo testing of 2CMO was carried out in \citet{jaeggli11}, and showed that for a photometric accuracy of $10^{-3}$ the magnetic field strengths greater than 1000 G can be fit to an accuracy of less than 10 G, and the field azimuth and inclination can be determined to less than $1^\circ$ for fits of the Fe I line pairs at either 6302 or 15650 \AA.  However the simulated lines had perfect symmetry.

\subsection{Results}
\begin{figure*}
	\begin{center}
		\includegraphics[width=7in]{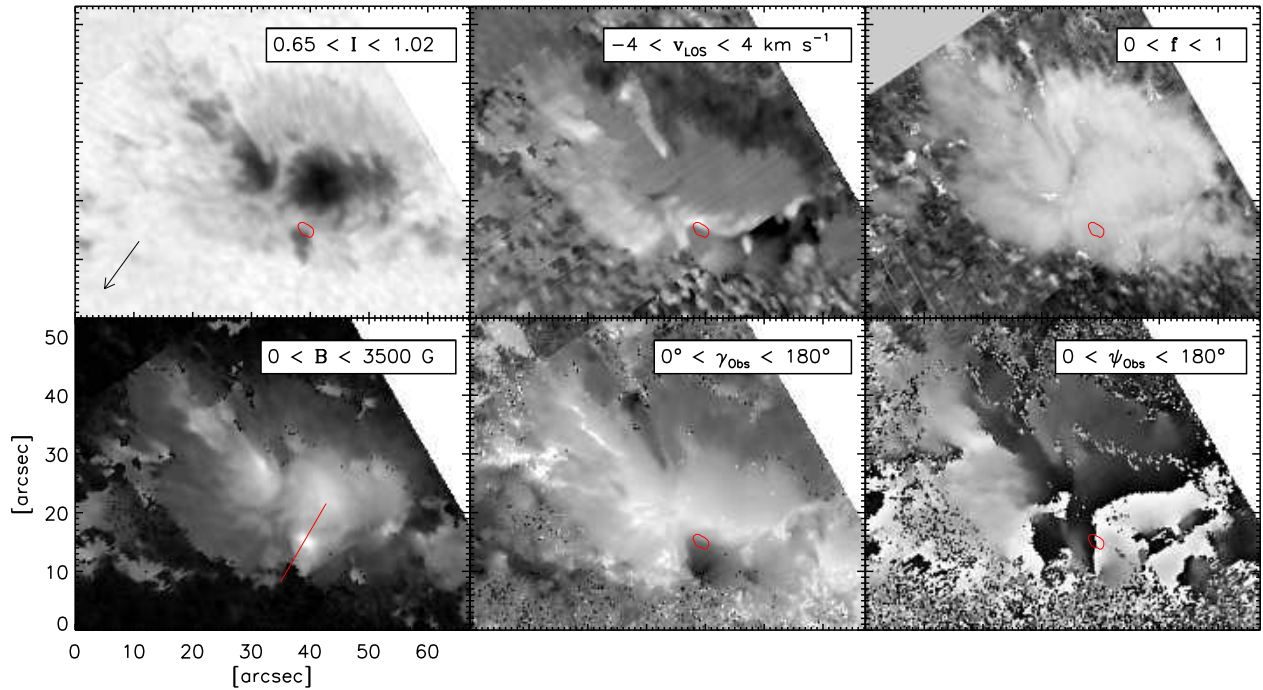}
	\end{center}
	\caption{Parameter maps from a sub-field of the FIRS 15650 \AA\ raster observation of NOAA 11035 taken 13:44:51 - 14:08:46 UT on Dec 17, 2009.  Continuum intensity (top left) was determined directly.  The other parameters are the result of inversion with the 2CMO 1M+SL model: line of sight velocity (top center), magnetic fill fraction (top right), magnetic field strength (bottom left), magnetic field inclination to the line of sight (bottom center), and the magnetic field azimuth in the plane of the sky (bottom right).}
	\label{fig:FIRS_maps}
\end{figure*}

Figures \ref{fig:SOTSP_maps} and \ref{fig:FIRS_maps} show maps resulting from the 1M+SL inversion of SOT/SP and FIRS data respectively.  Each figure shows the observed continuum intensity, the fitted Doppler velocity, magnetic fill fraction, magnetic field strength, magnetic field inclination, and magnetic field azimuth in the plane of the sky.  The red contour encloses the region of interest where the magnetic field strength is larger than 3000 G.  The black arrow in the upper left panel shows the direction toward the center of the solar disk.

The $\delta$-spot shows remarkable features which are present in both the SOT/SP and FIRS parameter maps.  Large magnetic field strengths perpendicular to the line of sight are present at the PIL between the umbra of the $\delta$-spot and the main sunspot umbra.  This concentration of magnetic field overlaps the top edge of the umbra and penumbra of the main spot.  In the continuum intensity map it is bounded by bright lobes on each side.  The portion of the shared penumbra to the west (right) of the parasitic umbra clearly belongs to the parasitic polarity, not the main sunspot.  Interestingly, the magnetic and velocity structure in this portion of the penumbra is relatively smooth, while other portions of the main sunspot penumbra show highly structured, filamentary lanes.  The velocity structure at the boundary of the parasitic polarity shows large blue shifts (positive values in the velocity color scale) of up to ~4 km s$^{-1}$ along the line of sight on the (upper and left) edges which lead interaction with the main sunspot, while the trailing (right) edge shows even stronger redshifts.  The systematic velocities of the limb and disk-side penumbra of the main sunspot display velocities typical of the Evershed effect \citep{evershed10}.

Interesting features in velocity and magnetic field are also visible in the like-polarity umbra to the upper left that is in the process of merging with the main sunspot umbra at this time, but we reserve discussion of this region for a different time.

\section{Direct Inspection of Spectra}\label{sec:anal2}
The high velocity signature so close to the strong magnetic feature casts some doubt on the reality of the magnetic field measurement.  A Milne-Eddington model with a single magnetic component can produce only symmetric Stokes Q and U, and anti-symmetric Stokes V profiles, but anomalous Q, U, and V profiles often occur in sunspot penumbrae \citep{bellotrubio04}.  In this $\delta$-spot Stokes profiles from different regions along the line of sight could become blended within a single pixel and produce a signature which is interpreted by the inversion code as a large field strength.  To cast some light on this possibility it is necessary to investigate the spectra directly.

\begin{figure*}
	\begin{center}
		\includegraphics[width=5.5in]{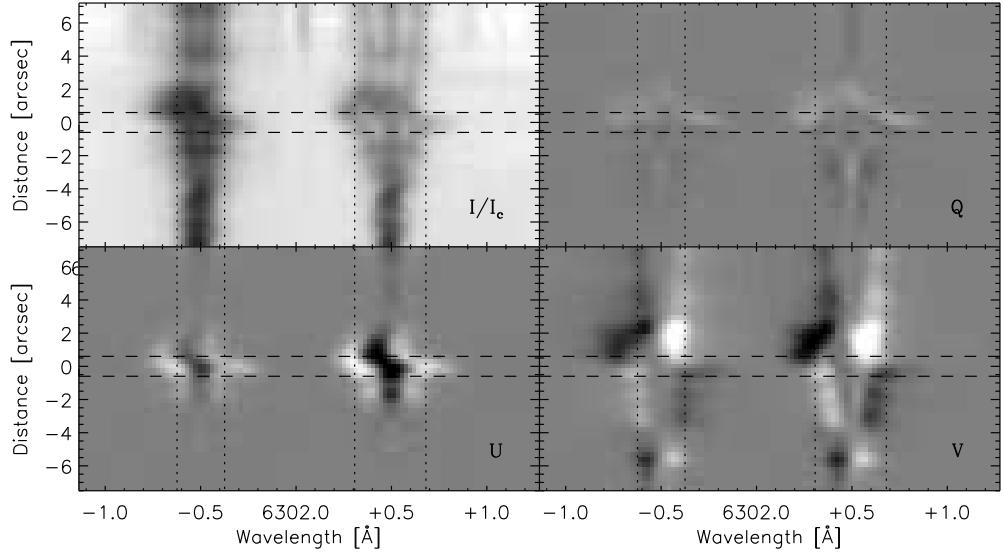}
	\end{center}
	\caption{Stokes spectra of the 6302 \AA\ region observed by the SOT/SP.  The spectra were extracted along a path indicated by the line in the continuum map of Figure \ref{fig:SOTSP_maps}.  The dotted vertical lines show the wavelength shift equivalent to a 4000 G field for each of the Fe I lines.  Individual profiles from the region enclosed by the horizontal dashed lines are shown in Figure \ref{fig:SOTSP_path_profile}.}
	\label{fig:SOTSP_path_spec}
\end{figure*}

\begin{figure*}
	\begin{center}
		\includegraphics[width=5.5in]{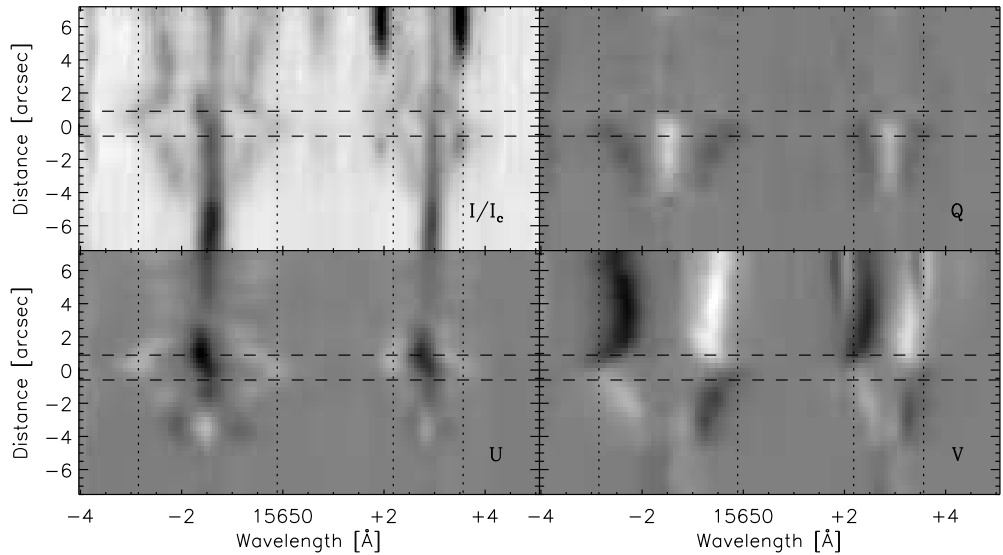}
	\end{center}
	\caption{Stokes spectra of the 15650 \AA\ region observed by the FIRS.  The spectra were extracted along a path indicated by the line in the continuum map of Figure \ref{fig:FIRS_maps}.  The dotted vertical lines show the wavelength shift equivalent to a 4000 G field for each of the Fe I lines.  Individual profiles from the region enclosed by the horizontal dashed lines are shown in Figure \ref{fig:FIRS_path_profile}.}
	\label{fig:FIRS_path_spec}
\end{figure*}

\begin{figure*}
	\begin{center}
		\includegraphics[width=7in]{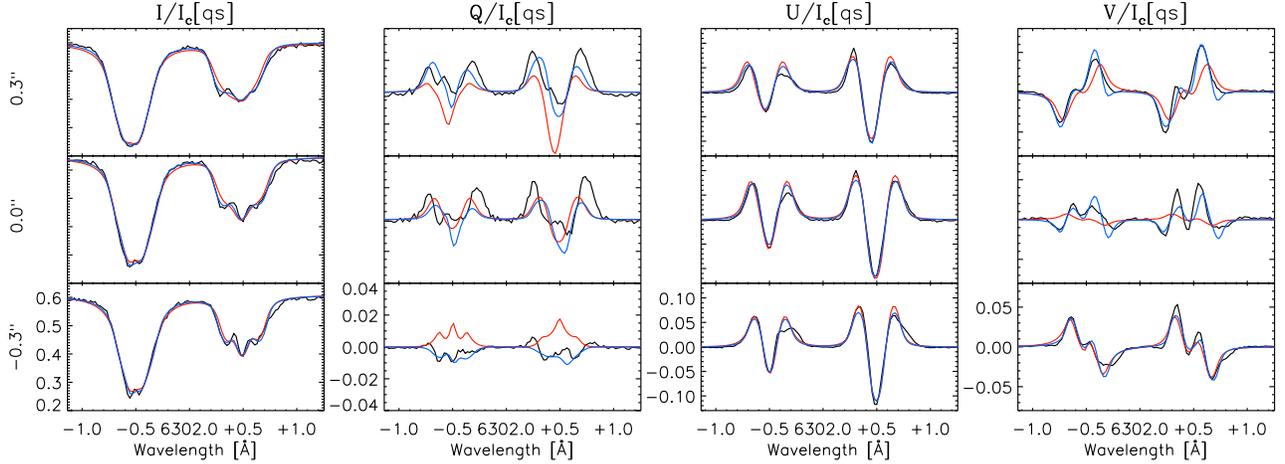}
	\end{center}
	\caption{Individual SOT/SP 6302 \AA\ Stokes profiles for the region enclosed by the dashed lines in Figure \ref{fig:SOTSP_path_spec}.  The black line shows the observed profile, the synthetic profiles from the 1M+SL and 2M inversion are shown in red and blue respectively.  The position of each profile is given at the far left}
	\label{fig:SOTSP_path_profile}
\end{figure*}

\begin{figure*}
	\begin{center}
		\includegraphics[width=7in]{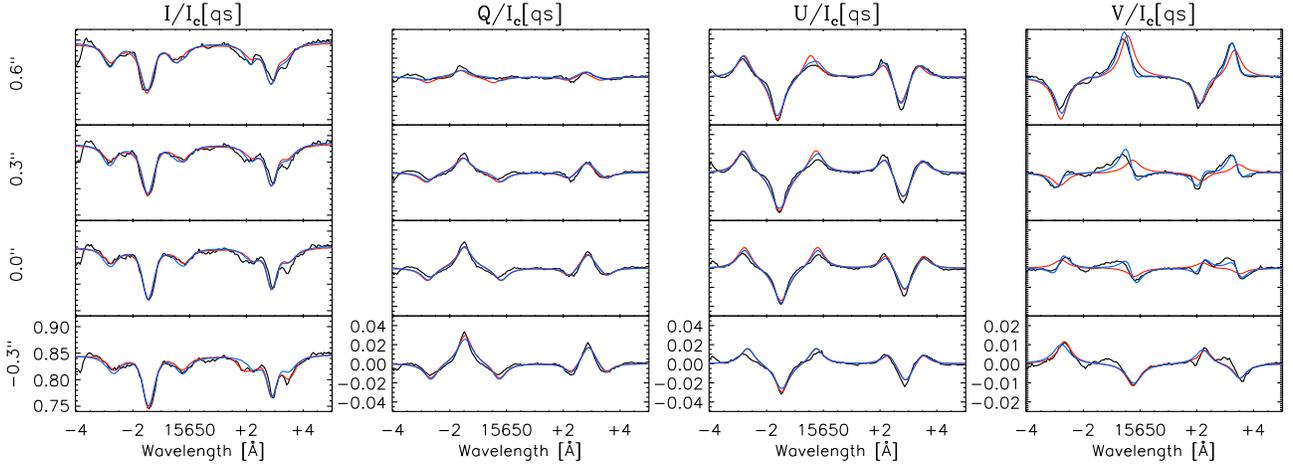}
	\end{center}
	\caption{Individual FIRS 15650 \AA\ Stokes profiles for the region enclosed by the dashed lines in Figure \ref{fig:FIRS_path_spec}.  The black line shows the observed profile, the synthetic profiles from the 1M+SL and 2M inversion are shown in red and blue respectively.  The position for each profile is given at the far left.}
	\label{fig:FIRS_path_profile}
\end{figure*}

Figures \ref{fig:SOTSP_path_spec} and \ref{fig:FIRS_path_spec} show the SOT/SP 6302 \AA\ and FIRS 15650 \AA\ spectra extracted by interpolating along a path crossing through the strongest field pixel approximately perpendicular to the magnetic inversion line and blue shifted region between the sunspot and the parasitic polarity (the red line in the lower left frame of Figures \ref{fig:SOTSP_maps} and \ref{fig:FIRS_maps}).  The Stokes I spectra shown in these figures have been divided by the continuum intensity at each position so that the line profiles can be seen clearly.  If unresolved blending of the profiles is happening, it will be most extreme across regions where the physical properties of the atmosphere change most rapidly.  The selected path crosses from the quiet-Sun, through the umbra of the parasitic polarity and across the inversion line into the main sunspot penumbra and umbra.  It is important to remember that the SOT/SP and FIRS spectra are from different times and may not be precisely aligned.  In Figures \ref{fig:SOTSP_path_spec} and \ref{fig:FIRS_path_spec}, the zero position on the y-axis indicates where the strongest field was measured.  The vertical dotted lines show the equivalent line positions for a 4000 G magnetic field for each of the Zeeman-split diagnostic lines. The individual Stokes profiles for the regions enclosed by the dashed lines are plotted in black in Figures \ref{fig:SOTSP_path_profile} (SOT/SP) and \ref{fig:FIRS_path_profile} (FIRS).

The SOT/SP spectra show a large change in velocity across the strong-field region.  The line core and wings are blue-shifted in the 2 arcsec region above the zero position, this is most clearly seen in the Stokes U component.  At the zero position, the line core is near the rest wavelength or very slightly redshifted, but the red wing is much broader than the blue wing.  The wing of the Fe I 6301.5 \AA\ line extends further relative to the 4000 G nominal splitting shown by the dotted lines than the wing of the Fe I 6302.5 \AA\ line.  Because these lines have different Land\'e g factors (1.5 and 2.5 respectively), this extension of the line wing must be due to a large redshift.

The FIRS Stokes Q and U components have been rotated to match the N-S slit orientation of the SOT/SP.  The blue-shifted line component seen in the SOT/SP spectra is also visible here.  The velocity shift is so large that the blueshifted line core is resolvable from the rest wavelength component, which is unpolarized and therefore must be due to an unresolved quiet-Sun component or, more likely, instrumental stray light.  At and below the zero position a slight redshift can be seen in the line core, but there is no extension in the red component of the line wing equivalent to the behavior seen in the Fe I 6302 \AA\ lines.  The longer wavelength and larger Land\'e g factor of the 15648.5 \AA\ line make it roughly three times more sensitive to magnetic field than the 6302 \AA\ line.  Magnetic field measurements with this line are less subject to velocity effects.  The clear splitting of the 15650 \AA\ lines in all four Stokes components leaves little doubt that the magnetic field is very large in this region.

\section{Inversion with 2 magnetic components}\label{sec:anal3}
\subsection{Setup}
The magnetic field strengths that can be approximated by eye from the profiles shown in Figures \ref{fig:SOTSP_path_spec} and \ref{fig:FIRS_path_spec} seem large, but for the individual profiles in Figures \ref{fig:SOTSP_path_profile} and \ref{fig:FIRS_path_profile}, all of the polarized Stokes components show anomalous profiles with multiple peaks and lack symmetry in Stokes Q and U, and anti-symmetry in Stokes V.  A Milne-Eddington inversion with a single magnetic component is not capable of fitting complex profiles, so the high field strength may arise from fitting the incorrect model to the data.

Complex profiles may result from two situations:  the blending of signals from different spatial regions with different velocity and magnetic field strengths and/or directions within a single resolution element, or gradients or discontinuities in the vertical dimension along the line forming region \citep{auer78}.  Gradients or discontinuities produce a net circular polarization (NCP) signal ($\int V d\lambda$), while unresolved blending of components should not show a signal if the NCP of each individual component is zero.

\begin{figure*}
	\begin{center}
		\includegraphics[width=7in]{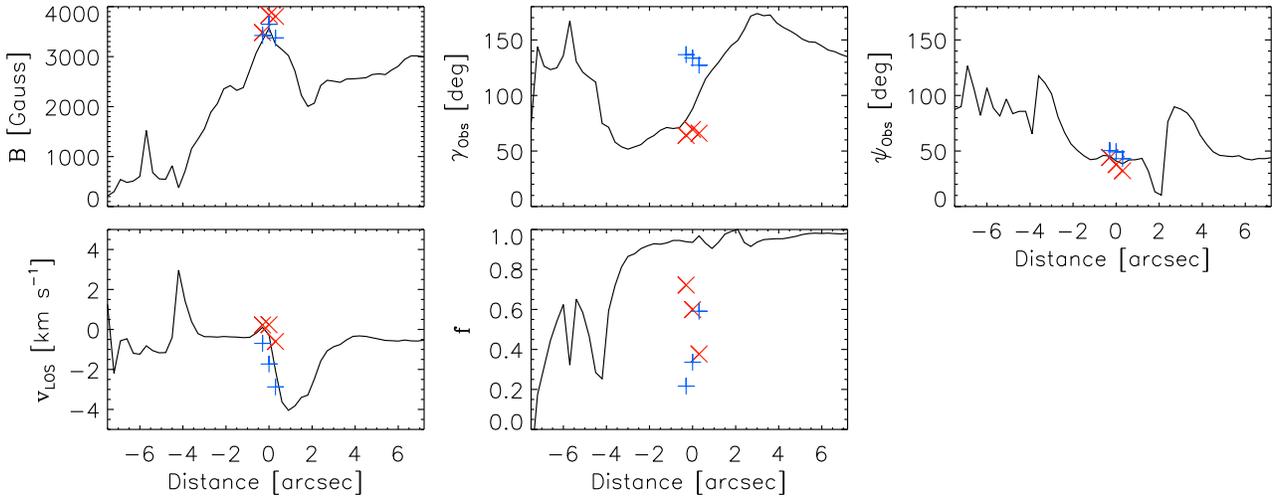}
	\end{center}
	\caption{Plots of the parameters resulting from the inversion of the SOT/SP 6302 \AA\ Stokes spectra shown in Figure \ref{fig:SOTSP_path_spec}.  The results from the 1M+SL inversion are shown by the solid black lines.  The results from the 2M inversion of the selected profiles in Figure \ref{fig:SOTSP_path_profile} are shown by the colored points, where the more red-shifted magnetic component is shown by a $\times$ and the more blue-shifted magnetic component is shown by a +.}
	\label{fig:SOTSP_path_param}
\end{figure*}

The NCP signal in the region of interest is much lower than in other regions of the penumbra, so it is assumed that the complexity in Stokes V is produced by unresolved blending of profiles with different magnetic field vectors and velocities.  This makes it possible to use a Milne-Eddington inversion with two magnetic components.  Inversions with two components have been used to interpret the more complex Stokes profiles that can be found in the penumbra where it is believed that there is a multi-height, uncombed structure with different magnetic field strengths, inclinations, and velocities arranged in narrow fibrils \citep{bellotrubio04}.  Two component inversions have also been previously employed in the inversion of $\delta$-spot spectra which show high velocities \citep{martinez94, lites02}.  The increased number of parameters in a multi-component inversion make it more likely that degeneracies will exist between the parameters, therefore it is important that the Stokes profiles display sufficient complexity.  A detailed study of degeneracies and errors in two component inversions is warranted but beyond the scope of the current work.  For the analysis of this region, with field strengths which appear to be in excess of 3000 G and velocities of order a few km s$^{-1}$, the adoption of a two component model seems more reasonable than for standard field strengths and velocities found in the penumbra.

A Milne-Eddington model with two partially independent magnetic components (hereafter called 2M) was used to invert a subset of the profiles that show complex behavior.  The components were sorted by their velocity and are labeled $r$ and $b$ for the more red or more blue shifted component.  The two magnetic components have independent parameters for magnetic field strength ($B_{r,b}$), azimuth ($\psi_{r,b}$), inclination ($\gamma_{r,b}$), and the Doppler velocity ($v_{r,b}$), but they share parameters for source function ($B_0$), source function gradient ($B_1$), Doppler width ($\sigma$), damping parameter ($\alpha$), and line to continuum absorption ratio ($\eta_0$).  An additional factor controls the relative fill fraction of the two magnetic components ($f_r$, $f_b=1-f_r$).

The 2M inversion was initialized by first performing the 1M+SL inversion of all the spectra shown in Figures \ref{fig:SOTSP_path_spec} and \ref{fig:FIRS_path_spec}.  The initial guess for the 2M inversion was provided by the neighboring regions on each side of the polarity inversion line with an appropriate filling factor.  A static stray light/non-magnetic contribution was estimated based on the umbral fill fraction in the 1M+SL inversion and this fraction was subtracted from the Stokes I profile prior to the inversion.  The same atomic parameters were used, but the molecular component was not included in the fitting for the FIRS 15650 \AA\ spectra for simplicity and due to the fact that they have a very weak intensity and polarization signature over the region of interest.

\subsection{Results}
\begin{figure*}
	\begin{center}
		\includegraphics[width=7in]{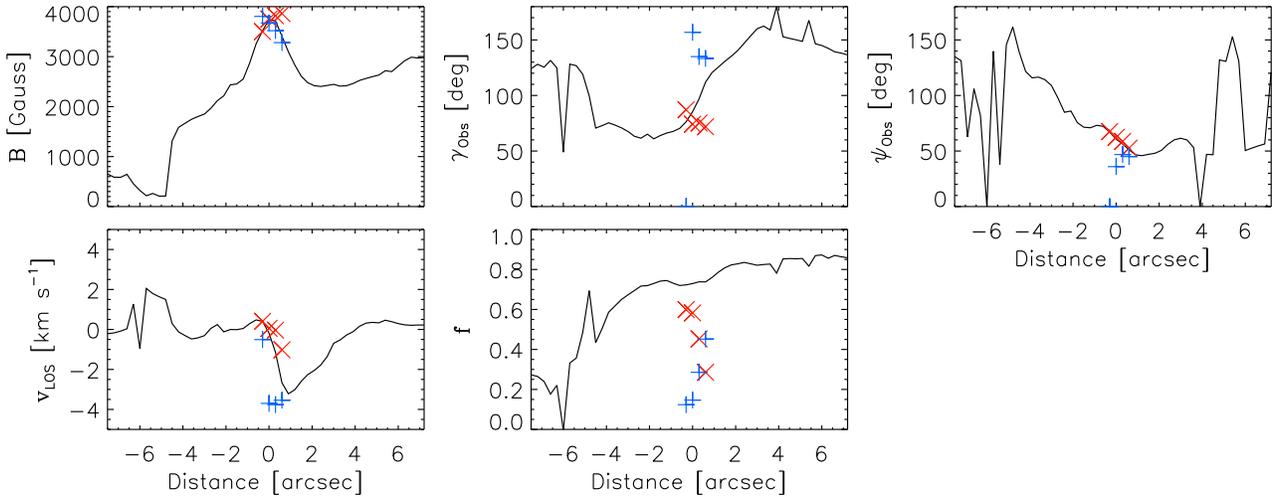}
	\end{center}
	\caption{Plots of the parameters resulting from the inversion of the FIRS 15650 \AA\ Stokes spectra shown in Figure \ref{fig:FIRS_path_spec}.  The results from the 1M+SL inversion are shown by the solid black lines.  The results from the 2M inversion of the selected profiles in Figure \ref{fig:FIRS_path_profile} are shown by the colored points, where the more red-shifted magnetic component is shown by a $\times$ and the more blue-shifted magnetic component is shown by a +.}
	\label{fig:FIRS_path_param}
\end{figure*}

The synthetic profiles produced by the 1M+SL (red line) and 2M (blue line) inversions of the SOT/SP and FIRS profiles in the regions of interest are plotted in Figures \ref{fig:SOTSP_path_profile} and \ref{fig:FIRS_path_profile} with the observed spectra (black line).  The stray light profile has been added back to the Stokes I synthetic spectrum from the 2M inversion for comparison.  The resulting parameters for both wavelengths from the one component inversion are shown by the solid black line in Figures \ref{fig:SOTSP_path_param} and \ref{fig:FIRS_path_param}.  The parameters selected by the two component  inversion are indicated by the plotted points.  The more red shifted component is indicated by red ($\times$)s and the more blue shifted component is indicated by blue (+)s.  The magnetic parameters are summarized in Table \ref{tab:2cmoresults}.

The fitted profiles from the initial inversion with the 1M+SL model do not do a good job fitting the well defined red and blue wings of Stokes I, and fail to fit the complex Stokes Q profile in the 6302 \AA\ spectra.  The complex Stokes V profiles are very poorly fit in both the visible and infrared spectra.  There is a mismatch between the apparent centers of the Stokes profiles which cannot be accounted for by a single magnetic component.  The 2M fits show great improvement for the Stokes I and V profiles, but the shape of Stokes Q is still not well fit.  The 2M synthetic profiles show an impressive, but imperfect, match to the asymmetries in the Stokes V profile.  The observed profiles are still more complex than two magnetic components are capable of describing.

The parameters produced by the 2M inversion show convincing, systematic behavior which is similar for the SOT/SP and FIRS results.  The 2M inversion selects a stronger magnetic component that is more red shifted or nearly at the rest velocity and shows a magnetic inclination more toward the observer.  The other component is slightly weaker by 100-500 G, more blueshifted, and inclined away from the observer.  The magnetic field azimuth angle for both components is approximately the same and close to the result from the 1M+SL inversion.  The field strengths of both magnetic components resulting from the 2M inversion of the SOT/SP data are higher than the field strength derived by the 1M+SL inversion, while the field strengths from the 2M inversion of the FIRS data are distributed higher and lower than the 1M result.  This indicates that a one component inversion may actually underestimate the average field strength when the Fe I 6302 \AA\ lines have complex profiles.

\begin{deluxetable*}{rrrrrrrrrr}
	\tablecaption{Magnetic vectors from the 2CMO inversions\label{tab:2cmoresults}}
	\tablehead{ & \multicolumn{4}{c}{1M+SL Inversion} & \multicolumn{4}{c}{2M Inversion} \\
	\hline
\colhead{Wavelength} & \colhead{Position} & \colhead{B} & \colhead{$\gamma_{Obs}$} & \colhead{$\psi_{Obs}$} & \colhead{$v_{LOS}$} & \colhead{B} & \colhead{$\gamma_{Obs}$} & \colhead{$\psi_{Obs}$} & \colhead{$v_{LOS}$} \\
		\colhead{[\AA]} & \colhead{['']}& \colhead{[G]} & \colhead{[deg]} & \colhead{[deg]} & \colhead{[km s$^{-1}$]} & \colhead{[G]} & \colhead{[deg]} & \colhead{[deg]} & \colhead{[km s$^{-1}$]}}
		\startdata
6302 & 0.3 & 3249. & 101.9 & 38.5 & -2.06 & 3804. & 65.9 & 32.2 & -0.60 \\
& & & & & & 3372. & 127.2 & 43.1 & -2.88 \\
\hline
& 0.0 & 3568. & 88.0 & 41.0 & -0.28 & 3832. & 69.5 & 37.8 & 0.24 \\
& & & & & & 3647. & 133.6 & 49.3 & -1.73 \\
\hline
& -0.3 & 3323. & 78.2 & 45.6 & 0.11 & 3482. & 64.1 & 44.2 & 0.22 \\
& & & & & & 3422. & 136.8 & 50.4 & -0.70 \\
\hline\hline
15650 & 0.6 & 3407. & 112.0 & 51.1 & -2.67 & 3861. & 71.9 & 52.6 & -1.02 \\
& & & & & & 3280. & 133.3 & 45.0 & -3.54 \\
\hline
& 0.3 & 3713. & 97.0 & 56.3 & -1.14 & 3815. & 75.2 & 58.7 & -0.04 \\
& & & & & & 3518. & 135.0 & 46.8 & -3.76 \\
\hline
& 0.0 & 3742. & 85.4 & 60.8 & -0.19 & 3751. & 74.1 & 62.2 & 0.05 \\
& & & & & & 3673. & 156.8 & 35.9 & -3.70 \\
\hline
& -0.3 & 3515. & 76.3 & 67.1 & 0.40 & 3500. & 87.2 & 67.6 & 0.41\\
& & & & & & 3805. & 0.3 & 0.0 & -0.51
		\enddata
\end{deluxetable*}

The infrared profile nearest the bottom in Figure \ref{fig:FIRS_path_profile} shows very similar results for the 1M+SL and 2M models, although the Stokes U and V profiles show obvious broadening in the red wings.  In Figure \ref{fig:FIRS_path_param} we can see that the filling factor of the blue-shifted component is about 10\% and the other parameters show outlying values relative to the others.  Because of the low filling factor this component does not contribute much to the overall shape of the profile (i.e. a single component produces the best fit), and the results from this point should be disregarded.

\section{Ambiguity Resolution}
Because of the symmetry of Stokes Q and U profiles, it is not possible to tell from the spectral profile which direction the magnetic field transverse to the line of sight is pointed.  There exist a variety of methods for resolving the problem of the magnetic field $180^\circ$ ambiguity \citep{metcalf06}.  The magnetic field of the lead sunspot in NOAA 11035 is complex, making accurate disambiguation with an automated technique difficult.  Nonetheless, disambiguation is necessary in order to retrieve the magnetic field vector relative to the orientation of the solar surface because the active region was observed a large distance from disk center at a heliocentric distance of $41.0^\circ$.

The magnetic field in the parasitic umbra should be somewhat radial.  This assumption was used to enforce a direction for the transverse magnetic field.  The magnetic field vector is then transformed into the local solar reference frame.  This technique is only applicable to the field immediately around the parasitic umbra, because this assumption of radial symmetry does not hold elsewhere in the active region.

\begin{figure*}
	\begin{center}
		\includegraphics[width=7in]{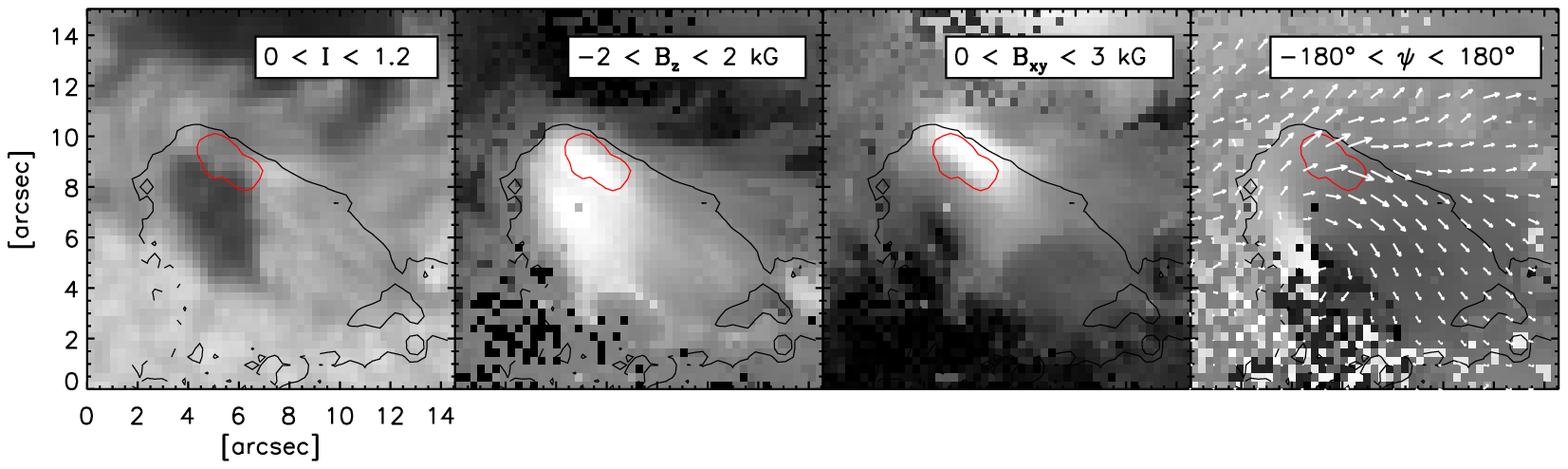}
	\end{center}
	\caption{Maps of the ambiguity-resolved magnetic field based on the SOT/SP inversion results, the panels from left to right show:  intensity, vertical and horizontal components of the magnetic field, and azimuth angle of the horizontal component.  Arrows on the right frame show the strength and direction of the horizontal field component.  The red contour shows the region where the total magnetic field strength is larger than 3000 G.  The black contour shows the magnetic polarity inversion line for the solar coordinate frame.}
	\label{fig:ambig_maps}
\end{figure*}

Only the 1M+SL inversion results from the SOT/SP are ambiguity resolved in a small region around the parasitic polarity.  The magnetic field results from FIRS are similar and of lower resolution.  Doing a disambiguation of the 2M results does not make sense because the disambiguation relies on continuous data and only a few datapoints have been inverted with the two component model, but it is possible to interpret the direction of the field based on the 1M ambiguity-resolved data.  The resulting maps of the vertical and horizontal components of the magnetic field, and the horizontal azimuth are shown in Figure \ref{fig:ambig_maps}.  The magnetic neutral line for the ambiguity-resolved field is plotted in black and the red contour shows where the total field is greater than 3000 G in the region of interest.  The size and directions of the arrows on the map of magnetic azimuth show the magnitude and direction of the horizontal field component.

A few interesting things are apparent in the maps.  The horizontal component of the magnetic field shows a large intensification relative to its surroundings.  The intensification in the vertical field component is not as pronounced and does not seem unusual in the context of the pore magnetic field.  The largest magnetic field is reached where the horizontal and vertical components are strong, making the field diagonal.  The magnetic field azimuth is nearly perpendicular to the PIL where the field is most intense (low shear), but in other regions along the PIL the azimuth is parallel to it (high shear).  The most vertical field in the parasitic umbra is not centered on the umbra as seen in the intensity image, but offset to the southeast, away from the main sunspot.

\begin{figure*}
	\begin{center}
		\includegraphics[width=4.5in]{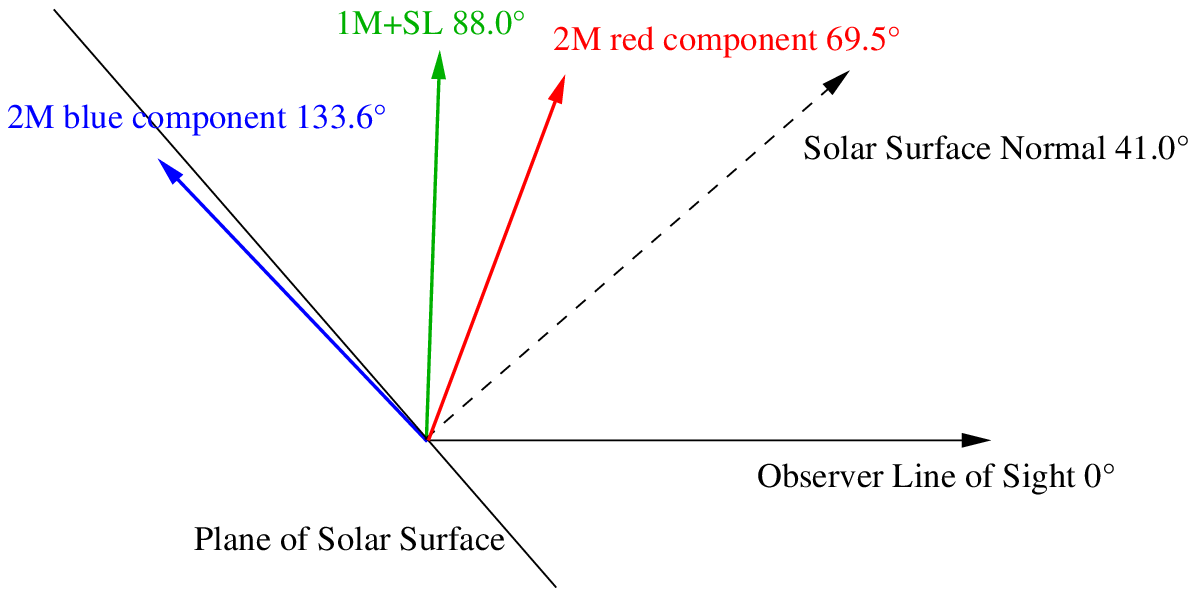}
	\end{center}
	\caption{A cartoon demonstrating the geometry of the magnetic field resulting from the 2M inversion in the solar coordinate frame.}
	\label{fig:geom_cartoon}
\end{figure*}

The magnetic field azimuth direction in the strong field region is very close to the heliocentric position angle of the observation, and therefore lies near the projected line of sight.  This makes interpretation of the geometry for the two magnetic components simple because it is possible to use just the inclination angle when considering projection effects.  Figure \ref{fig:geom_cartoon} shows a cartoon of the magnetic field inclination angles in relation to the observed line of sight and the local solar coordinate frame for the center position from the SOT/SP inversions.  The 2M blue component lies almost horizontal to the surface, while the 2M red component is closer to the vertical direction.  The inclination angle inferred from the 1M+SL inversion is diagonal and nearly 90$^\circ$ to the line of sight.

\section{Discussion and Conclusions}\label{sec:discuss}
This work has investigated an unusually strong magnetic feature that appeared in a $\delta$-spot between the parasitic and main polarities in NOAA 11035 and was observed over the course of two days.  The initial analysis with a Milne-Eddington inversion showed a patch of very strong, magnetic field, transverse to the line of sight with a peak strength of $\sim3700$ G seen in the infrared.  This unusually high field strength was called into question due to large velocities, the coincidence of the polarity inversion line in the observed frame, and complex profiles seen in the Stokes Q, U, and V spectral profiles.  Select complex spectra were inverted using a Milne-Eddington model with two magnetic components and it was found that the feature is well described.  In general, the more blueshifted component shows velocities of a few km s$^{-1}$ toward the observer and a slightly weaker magnetic field which is nearly horizontal in the local solar reference frame.  The more redshifted component has velocities of a fraction of a km s$^{-1}$ and a slightly stronger magnetic field which is oriented more vertically.  Similar results were seen in both the Fe I magnetic field diagnostics taken with SOT/SP at 6302 \AA\ onboard Hinode and with FIRS at 15650 \AA\ on the DST.  The azimuth vector of the magnetic field in the solar coordinate frame is not sheared at the PIL where the magnetic field is strongest.

These observations are unusual even among the few observations of strong fields in $\delta$-spots previously published.  The magnetic field measured at the polarity inversion line of NOAA 11035 is as strong as those reported by \citet{tanaka91} and \citet{zirin93a,zirin93b}, but the active region is smaller and far less complex than the sunspot groups analyzed in those papers.  Recent close analysis of a $\delta$-spot magnetic field by \citet{cristaldi14} does show a small intensification of the transverse field at the PIL up to 2300 G, but the analysis of another $\delta$-spot by \citet{balthasar14} does not show any intensification.

This strong magnetic feature lasts for two days, much longer than the 20 minute dynamical timescale of the photosphere.  The presence of this very strong, nearly horizontal magnetic field does not make sense in the context of the standard model of sunspot magnetohydrostatic equilibrium.  In general, sunspot magnetic fields are strongest and vertical in the umbra and weaker and more horizontal in the penumbra of a sunspot.  Very strong horizontal fields are generally not permitted because they are buoyantly unstable \citep{parker55}, and this mechanism is believed to drive the rise of active region magnetic flux from the solar interior through the convection zone and into the photosphere where the field lines closely approximate a potential field.  In order to keep this strong field region at the surface of the photosphere, there must be strong magnetic field above it that prevents it from rising further, or some mechanism beneath the photosphere that prevents the flux from rising.

Another mechanism could explain the strong field seen at the PIL.  Magnetic fields of up to 7000 G have been inferred from spectropolarimetric observations of high-velocity penumbral downflows based on Fourier interpolation and inversion of the observed Stokes profiles \citep{vannoort13}.  These features are far below the resolution limit and originate at great depth in the solar surface which is visible because of the evacuation of material within the downflow.  If the interior of the parasitic polarity umbra was very cool, then it might have been possible to observe a higher field strength due to decreased opacity (the Wilson Depression).  If this were the case, strong OH lines would be seen where the field is strongest.  However, the strong field region shows very weak OH lines and also seems to be related to penumbral structure which is higher up than the typical umbral photosphere.

It is possible to explain the observations without invoking opacity effects if the sunspot canopy field is laying over the top of the small parasitic polarity region, confining the polarity and leading to a compression of the field on one side.  Three pieces of evidence indicate this is happening:
\begin{enumerate}
\item The magnetic and velocity structure inferred from an inversion with two magnetic components shows that the components are roughly equal in strength, but one component is orientated more vertically while the other is more horizontal with respect to the solar surface.  The velocity along the horizontal component is higher and directed out of the sunspot while the more vertical component is closer to the rest velocity and more horizontal.  This structuring of the magnetic field is similar to the two component picture of the penumbra of flux tubes imbedded in a background field presented in \citet{bellotrubio04}.
\item The velocity at the interface between the parasitic polarity and the main sunspot is structured along penumbral fibrils, indicating that it is related to the Evershed flow of the penumbra.  The velocity enhancement in this section of the penumbra relative to other areas of the sunspot penumbra may be because the flow is more directly along the line of sight in this region due to the altered magnetic structure.
\item Inside the parasitic umbra, the lowest (most vertical) inclination of the magnetic field relative to the solar surface is offset from the center of the umbra toward the southeast, directly away from the host sunspot.  This region is well described by a single magnetic component, so it is certain that the field is pointed this way. 
\end{enumerate}

It is mysterious why this magnetic configuration is permitted, and why the field does not simply reconnect and vanish or the opposite polarities push apart to reduce the magnetic tension that must be very high at the PIL.  The footpoints of the magnetic field reaching down into the convection zone may be confined.  Paper II will look in depth at the evolution and structure in the chromosphere and corona to explore how this feature was produced and what it tells us about magnetic reconnection and penumbral processes.

\acknowledgments
We dedicate this work to the memory of DST observer Mike Bradford, without whom these observations would not have been possible.

The author thanks H. Lin, P. Judge, C. Kankelborg, A. Daw, D. Rabin, and L. Tarr for useful discussion and encouragement, K. Yoshimura for assistance with image coalignment, and the anonymous referee for helpful comments on this work.

This research was supported by an appointment to the NASA Postdoctoral Program at Goddard Space Flight Center administered by Oak Ridge Associated Universities through a contract with NASA.

Hinode is a Japanese mission developed and launched by ISAS/JAXA, with NAOJ as domestic partner and NASA and STFC (UK) as international partners. It is operated by these agencies in co-operation with ESA and NSC (Norway).

{\it Facilities:} \facility{Dunn}, \facility{Hinode}.


\begin{thebibliography}{}
	\bibitem[{Auer, Heasley, \& House}(1977)]{auer77}
	{Auer}, L.~H., {Heasley}, J.~N., \& House, L.~L. 1977, \solphys, 55, 47

	\bibitem[{Auer \& Heasley}(1978)]{auer78}
	{Auer}, L.~H. \& {Heasley}, J.~N. 1978, \aap, 64, 67

	\bibitem[{Balthasar et al.}(2014)]{balthasar14}
	{Balthasar}, H., {Beck}, C., {Louis}, R.~E., {Verma}, M., \& {Denker}, C. 2014, ASPC, 489, 39

	\bibitem[{Bellot Rubio, Balthasar, \& Collados}(2004)]{bellotrubio04}
	{Bellot Rubio}, L.~R., {Balthasar}, H., \& {Collados}, M. 2004, \aap, 427, 319

	\bibitem[{Berdyugina \& Solanki}(2002)]{berdyugina02}
	{Berdyugina}, S.~V. \& {Solanki}, S.~K. 2002, \aap, 385, 701

	\bibitem[{Borrero et al}(2003)]{borrero03}
	{Borrero}, J.~M., {Bellot Rubio}, L.~R., {Barklem}, P.~S., {del Toro Iniesta}, J.~C., 2003, \aap, 404, 749

	\bibitem[{Cristaldi et al}(2014)]{cristaldi14}
	{Cristaldi}, A., {Guglielmino}, S.~L., {Zuccarello}, et al. 2014, \apj, 789, 162

	\bibitem[{Denker et al.}(2007)]{denker07}
	{Denker}, C., {Deng}, N., {Tritschler}, A., \& {Yurchyshyn}, V. 2007, \solphys, 245, 219

	\bibitem[{De Pontieu et al.}(2014)]{depontieu14}
	{De Pontieu}, B., {Title}, A.~M., {Lemen}, J.~R., et al. 2014, \solphys, 289, 2733
	
	\bibitem[{Elmore et al.}(1992)]{elmore92}
	{Elmore}, D.~F., {Lites}, B.~W., {Tomczyk}, et al. 1992, SPIE, 1746, 22

	\bibitem[{Evershed}(1910)]{evershed10}
	{Evershed}, J. 1910, \mnras, 78, 217

	\bibitem[{Guo, Lin, \& Deng}(2014)]{guo14}
	{Guo}, J., {Lin}, J., \& {Deng}, Y. 2014, \mnras, 441, 2208

	\bibitem[{Hale et al.}(1919)]{hale19}
	{Hale}, G.~E., {Ellerman}, F., {Nicholson}, S.~B., \& {Joy}, A.~H. 1919, \apj, 49, 153
	
	\bibitem[{Ichimoto et al.}(2008)]{ichimoto08}
	{Ichimoto}, K., {Lites}, B., {Elmore}, D., et al. 2008, \solphys, 249, 179

	\bibitem[{Jaeggli et al.}(2010)]{jaeggli10}
	{Jaeggli}, S.~A., {Lin}, H., {Mickey}, D.~L., et al. 2010, \memsai, 81, 763

	\bibitem[{Jaeggli}(2011)]{jaeggli11}
	{Jaeggli}, S. A. 2011, Ph.D.T.

	\bibitem[{Jaeggli, Lin, \& Uitenbroek}(2012)]{jaeggli12}
	{Jaeggli}, S. A., {Lin}, H., \& {Uitenbroek}, H. 2012, \apj, 745, 133

	\bibitem[{Jefferies, Lites, \& Skumanich}(1989)]{jefferies89}
	{Jefferies}, J., {Lites}, B.~W., \& {Skumanich}, A. 1989, \apj, 343, 920
	
	\bibitem[{Jiang et al.}(2012)]{jiang12}
	{Jiang}, Y., {Zheng}, R., {Yang}, J., et al. 2012, \apj, 744, 50
	
	\bibitem[{Kuhn et al.}(1994)]{kuhn94}
	{Kuhn}, J~.R., {Balasubramaniam}, K.~S., {Kopp}, et al. 1994, \solphys, 153, 143
	
	\bibitem[{Kunzel}(1960)]{kunzel60}
	{Kunzel}, H. 1960, AN, 285, 271
	
	\bibitem[{Kunzel}(1965)]{kunzel65}
	{Kunzel}, H. 1965, AN, 288, 177
	
	\bibitem[{Lites et al.}(2013)]{lites13a}
	{Lites}, B.~W., {Akin}, D.~L., {Card}, G., et al. 2013, \solphys, 283, 579
	

	\bibitem[{Lites, Socas-Navarro, \& Skumanich}(2002)]{lites02}
	{Lites}, B.~W., {Socas-Navarro}, H., \& {Skumanich}, A. 2002, \apj, 575, 1131
	
	\bibitem[{Lites et al.}(1995)]{lites95}
	{Lites}, B.~W., {Socas-Navarro}, H., {Skumanich}, A., \& {Shimizu}, T. 1994, \apj, 446, 877

	\bibitem[{Liu et al.}(2005)]{liu05}
	{Liu}, C., {Deng}, N., {Liu}, Y., et al. 2005, \apj, 622, 722
	
	\bibitem[{Livingston et al.}(2006)]{livingston06}
	{Livingston}, W., {Harvey}, J.~W., {Malanushenko}, O.~V, \& {Webster}, L. 2006, \solphys, 239, 41

	\bibitem[{Mart\'inez Pillet et al.}(1994)]{martinez94}
	{Mart\'inez Pillet}, V., {Lites}, B.~W., {Skumanich}, A., \& {Degenhardt}, D. 1994, \apj, 425, L113

	\bibitem[{Matta \& Reichel}(1971)]{matta71}
	{Matta}, F. \& {Reichel}, A. 1971, Math Comp, 25, 339

	\bibitem[{Metcalf et al.}(2006)]{metcalf06}
	{Metcalf}, T.~R. et al. 2006, SolPhys, 237, 267

	\bibitem[{Nave et al.}(1994)]{nave94}
	{Nave}, G.,  {Johansson}, S., {Learner}, R.~C.~M., {Thorne}, A.~P., \& {Brault}, J.~W. 1994, \apjs, 94, 221

	\bibitem[{Orozco Suarez et al.}(2010)]{orozcosuarez10}
	{Orozco Suarez}, D. et al. 2010, \aap, 518, 2

	\bibitem[{Parker}(1955)]{parker55}
	{Parker}, E.~N. 1955, \apj, 121, 491

	\bibitem[{Rees \& Semel}(1979)]{rees79}
	{Rees}, D.~E. \& {Semel}, M.~D. 1979, \aap, 74, 1

	\bibitem[{Rimmele et al.}(2004)]{rimmele04}
	{Rimmele}, T., {Richards}, K., {Hegwer}, S., et al. 2004, SPIE, 5171, 179

	\bibitem[{Sammis, Tang, \& Zirin}(2000)]{sammis00}
	{Sammis}, I., {Tang}, F., \& {Zirin}, H. 2000, \apj, 540, 583

	\bibitem[{Shimizu et al.}(2007)]{shimizu07}
	{Shimizu}, T. et al. 2007, \pasj, 59, 845
	
	\bibitem[{Skumanich \& Lites}(1987)]{skumanich87}
	{Skumanich}, A. \& {Lites}, B.~W. 1987, \apj, 322, 473
	
	\bibitem[{Spruit}(1974)]{spruit74}
	{Spruit}, H.~C. 1974, \solphys, 34, 277

	\bibitem[{Tanaka}(1991)]{tanaka91}
	{Tanaka}, K. 1991, \solphys, 136, 133

	\bibitem[{van Noort et al.}(2013)]{vannoort13}
	{van Noort}, M., {Lagg}, A., {Tiwari}, S.~K., \& {Solanki}, S.~K. 2013, \aap, 557, 24
	
	\bibitem[{Zirin \& Liggett}(1987)]{zirin87}
	{Zirin}, H. \& {Liggett}, M.~A. 1987, \solphys, 113, 267
	
	\bibitem[{Zirin \& Wang}(1993a)]{zirin93a}
	{Zirin}, H. \& {Wang}, H. 1993, \nat, 363, 426L
	
	\bibitem[{Zirin \& Wang}(1993b)]{zirin93b}
	{Zirin}, H. \& {Wang}, H. 1993, \solphys, 144, 37

\end{thebibliography}
\end{document}